\documentstyle[editedvolume,psfig]{crckapb}

\begin{opening}
\title{The pulsational behaviour of variable stars
in the lower part of the instability strip}

\author{Ennio PORETTI}
\institute{Osservatorio Astronomico di Brera\\
Via E. Bianchi, 46 -- 23807 Merate, Italy\\
E-mail: poretti@merate.mi.astro.it}

\end{opening}

\runningtitle{Lower Instability Strip Variable stars}

\begin{document}

%\keywords{}

\section{Introduction}
The lower part of the classical instability strip and the surrounding main
sequence are populated by a large variety of pulsating variable stars. In
the past years a great effort was made by the stellar group
of Brera Observatory (E. Antonello, M. Bossi, 
L. Mantegazza, E. Poretti, F.M. Zerbi) to collect a large amount of
photometric and spectroscopic data on  $\delta$ Sct stars, the more
numerous class.
Observations were carried out
using the telescopes located in Merate (0.50 m and 1.02 m; a third telescope
having a diameter of 1.32 m will be available in the near future)
and in La Silla (European Southern Observatory, Chile; we
used the 0.50~m, 0.90~m, 1.0~m to make the photometric observations, the
1.4-m CAT for the spectroscopic ones). 

In this paper I will describe our approach to the study of $\delta$
Sct stars and some results we obtained; the reader can find other
related aspects in the
contributions written by F.M.~Zerbi in this volume. As a first step, in
Sect. 2.1 I show as the light curves of high--amplitude $\delta$ Sct stars 
can be described in a synthetic and powerful way by using the Fourier
decomposition. In Sect.~2.2 some interesting cases are presented, i.e. the
peculiar light curves of some double--mode  pulsators and the
multiperiodicity of V974 Oph. As regards the small amplitude $\delta$
Sct stars, the results which can be extracted from the photometry and
(high-resolution) spectroscopy are discussed in Sect.~3; the importance
of combining the two techniques into the synergic approach is stressed.
As a final step, an excursus on the $\gamma$ Dor stars is presented in
Sect.~4; the rich pulsational content makes these stars very similar
to $\delta$ Sct variables.

The $\delta$ Sct class (DSCT in the GCVS notation) contains now most of the 
stars previously classified as $\delta$ Sct, SX Phe, AI Vel, RRs variables.
The only separation now maintained is between Population I (i.e. DSCT stars) and
Population II (i.e. SXPHE stars). The amplitude of the light variation is
not considered a physical discriminant and hence both high (up to 0.50 mag) and
small (down to the mmag level) amplitude pulsators are included in the DSCT
class. For sake of clarity, let us consider separately the two subclasses.

\section{The high--amplitude $\delta$ Sct stars}

\subsection{The Fourier parameters}

Most of these variable stars are single--periodic, showing a light curves 
significatively deviating from a sine shape. In such a case it
is possible to fit the measurements by a sum of cosine functions having 
frequencies $f, 2f, 3f, ... $ (Fourier decomposition) and then to study the
particularities of the phase and amplitude parameters (see Poretti, this
volume, for the application of this technique to Cepheid
light curves).

As a natural extension of this work toward the lower part of the
instability
strip, Antonello et al. (1986) applied the Fourier
decomposition to the light curves of high amplitude ($>$0.20 mag), short
period ($P<$ 0.35 d) $\delta$ Sct stars, generally considered to be very
stable single--mode pulsators. Later, Poretti et al. (1990)
continued this study covering all the
available sample, actually constituted by 30 stars, i.e. all the
objects which could be observed by photoelectric photometry, plus 
a few stars observed with CCD detectors.
A more robust sample is necessary to
better investigate some unclear observational facts, still open even
considering the output of large--scale projects as MACHO and OGLE:

\begin{enumerate}
\item Poretti et al. (1990) suggested a bimodal distribution of the 
amplitude ratios $R_{21}=A_{2f}/A_f$. The matter is controversial since
contradictory results were obtained
by Templeton et al. (1998) and by Morgan et al. (1998).
From  a theoretical point of view, the mechanism which
limits the pulsational amplitude is far from being well understood and
Fourier plots can get new light on the subject;
\item  The progression of the phase differences $\phi_{21}=\phi_{2f}-2\phi{f}$
is well defined, but a change in the slope is observed at P=0.09~d.
The observational fact that no first--overtone pulsator
was found among high amplitude $\delta$ Sct stars (no sequence similar to
that of $s$-Cepheids was found in the $\phi_{21}-P$ plane)
need to be confirmed. We remind that overtone pulsators are
quite common in small amplitude $\delta$ Sct stars;
\item Poretti \& Antonello (1988) discovered some stars
showing a descending branch of the light curve steeper than the
ascending one
(V1719 Cyg variables). The implications on stellar models were
stressed by Antonello et al. (1988) and by Guzik (1992), but
the physical conditions necessary to generate such a light
curve are not yet well established (helium diffusion? resonances?).
In the next section the V1719 Cyg light curves will be further
discussed.
\end{enumerate}
\begin{figure*}
\centerline{\psfig{figure=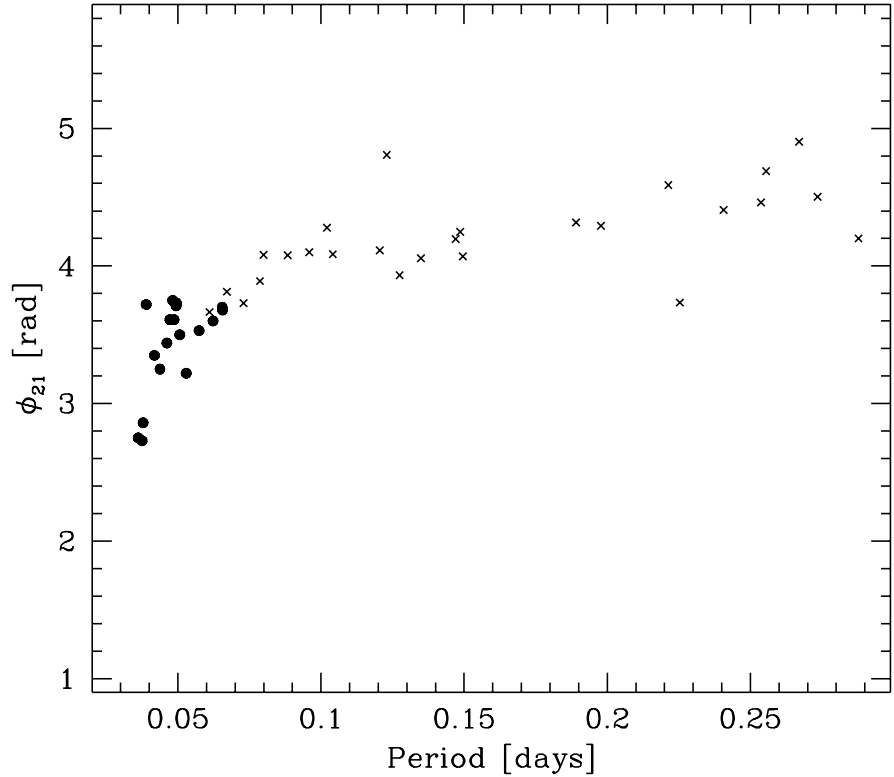,width=12.cm}}
\caption{The $\phi_{21}$ parameters as function of the period. Dots
are SX Phe stars in the globular cluster $\omega$ Cen, crosses are
Pop.~I and Pop.~II galactic variables.}
\end{figure*}

Recently, well-defined light curves of SX Phe stars belonging to the
globular cluster $\omega$ Centauri were decomposed (Poretti 1999). As
a result, an excellent overlap between globular and galactic stars
exists in the $\phi_{21}-P$ plane (Fig.~1); moreover, some
peculiar cases (an anomalous light curve and a group of slightly
deviating $\phi_{21}$ values) were found (see Poretti 1999 for details).

\subsection{Peculiar light curves}

\begin{figure*}
\centerline{\psfig{figure=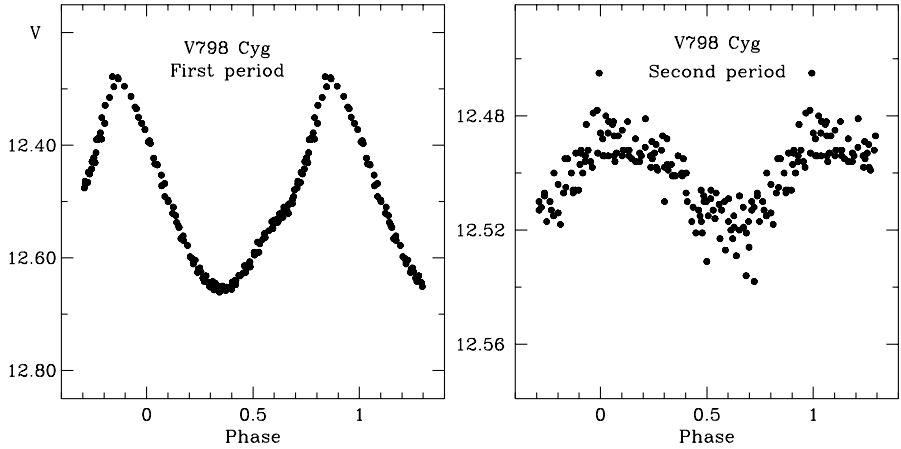,width=12.cm}}
\caption{CCD photometry of high amplitude $\delta$ Sct stars:
the two periods observed  in the light curve of V798 Cyg. Note the steeper
descending branch of the first period (left panel) and the small amplitude
of the second period (right panel).}
\end{figure*}

\begin{figure*}
\centerline{\psfig{figure=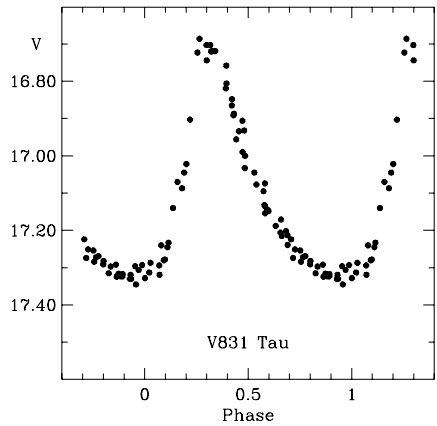,width=6.cm}}
\caption{CCD photometry of high amplitude $\delta$ Sct stars: the monoperiodic
variable V831 Tau shows a very large amplitude and a short period (0.0643 d).} 
\end{figure*}

\begin{figure*}
\centerline{\psfig{figure=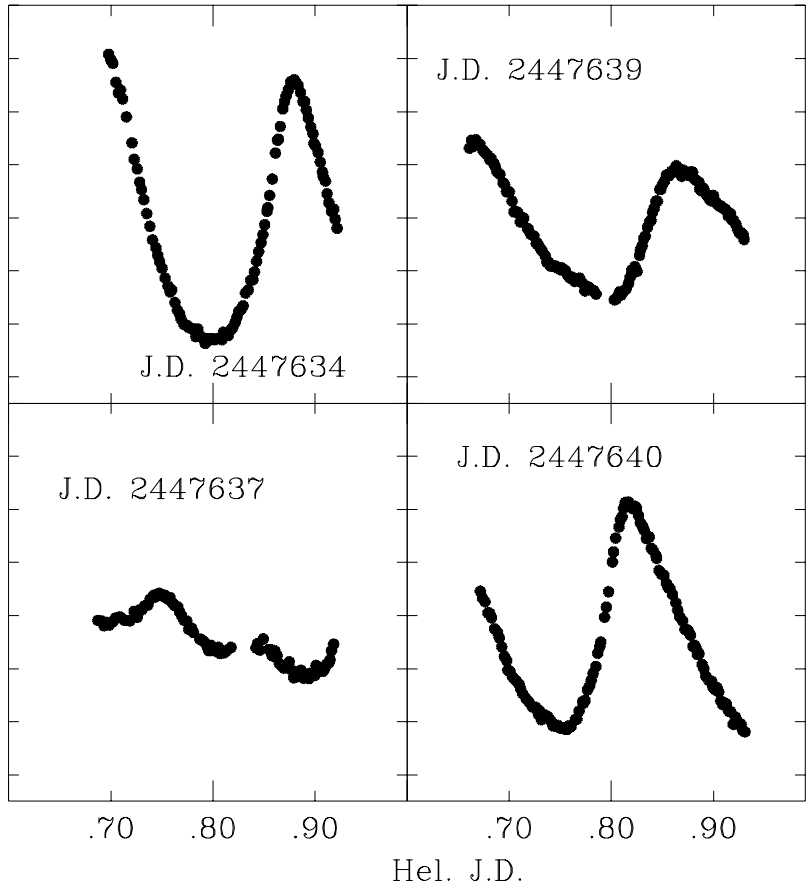,width=12.cm}}
\caption{The light curves of V974 Oph on four close nights: the strong
changes in the shape are the largest ever observed in amplitude in a 
$\delta$ Sct star. Ticks on the vertical axis are separated by 0.10 mag.}
\end{figure*}

Figure 2 shows the light curves of V798 Cyg: the first period (left panel) has 
a rising branch covering more than half period, owing to a bump clearly visible
just after the minimum light. This kind of asymmmetry is
quite uncommon in pulsating star light curves (Poretti \&
Antonello 1988). Respect with to preliminary results described there, it 
should be noted that new CCD measurements (Musazzi et al. 1998) confirm not
only the existence of a  second term, but also that its frequency is
6.41~c~d$^{-1}$. This means that $f_1/f_2$=0.80, as happens for the other star
showing the same asymmetrical light curve, i.e. V1719 Cyg. The possible 
link between the double--mode pulsation and asymmetrical light curve deserves
further attention in the future. 

In Fig.~3 the spectacular light curve of the 17th--mag star V831 Tau
is presented: the amplitude is about 0.70 mag and since the period is 
92 min, the ascending branch is only 20 min long. It reminds more known and
brighter stars as CY Aqr and DY Peg.

Among the high amplitude $\delta$ Sct stars, V974 Oph is an unique case.
Its multimode nature became evident after an observing run at the ESO 50--cm
telescope in April 1989; Fig.~4 shows an example of the dramatic changes
occurring over a short time baseline. Hence, multimode pulsation can be
acting also in high amplitude pulsators, not only in small amplitude
ones (V974 Oph can reach an amplitude of 0.5 mag in $B$ light).

\begin{figure*}
\centerline{\psfig{figure=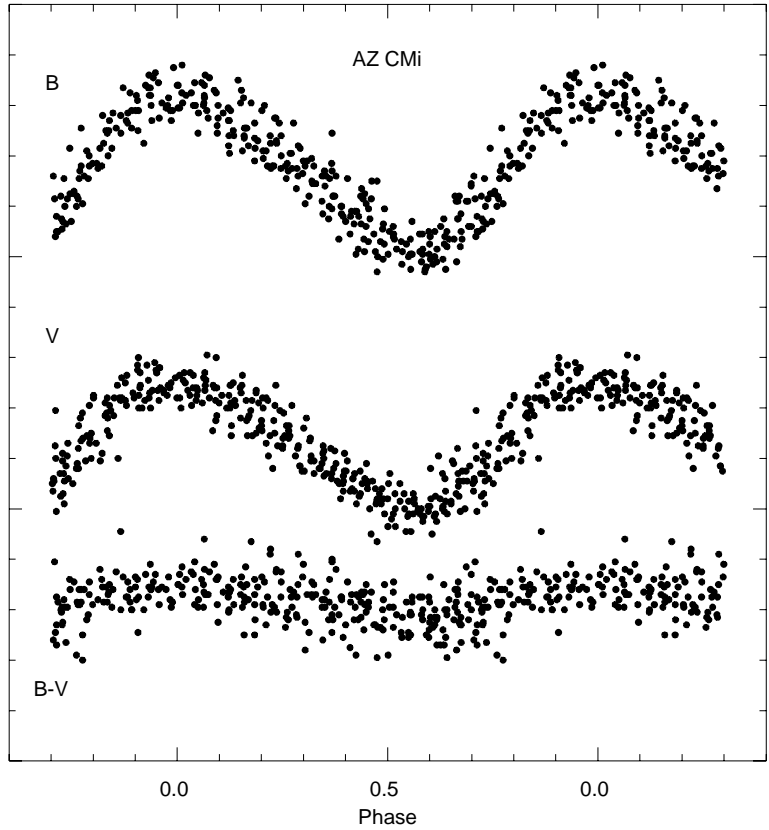,width=12.cm}}
\caption{Photoelectric measurements of the $\delta$ Sct star AZ CMi. The
light variation has a full amplitude of 0.07 mag only, but the curve is
very  regular.}
\end{figure*}
\section{The small amplitude $\delta$ Sct stars}

\subsection{The photometric approach}
The photometric and spectroscopic campaigns secured by our group (Tab.~1)
are proving that the presence of a rich
mixture of simultaneously excited radial and nonradial $p$-mode is quite
common among $\delta$ Sct stars. Such a characteristic makes them an ideal
tool for asteroseismologic studies. Moreover, since these stars cover a wide
range of evolutionary stages, ranging from pre-main-sequence to the giant
one, they offer the possibility to probe stellar interiors in very different
conditions.

In the small amplitude $\delta$ Sct stars we meet the reverse case respect with
to the high amplitude ones: multiperiodicity is quite common and 
monoperiodicity is rare. Figure~5 shows the light curve of AZ CMi, a
remarkable example of small amplitude, monoperiodic $\delta$ Sct star.
In the case of a monoperiodic star, also single--site measurements can
 provide a very useful check
of the constancy of the period (see Riboni et al. 1994 for the case of
$\beta$ Cas), but also the amplitude needs to be monitored (Poretti et al.
1996). As a matter of fact, the amplitudes of the excited modes are often
variable in time, as can be verified by comparing different observing
seasons on the same star.

\begin{table}
\caption{List of campaigns on $\delta$ Sct stars carried out by the 
stellar group of Brera Observatory. N is the number of photoelectric
measurements or of high--resolution spectrograms. Appropriate references
are quoted.}
\begin {tabular}{lccrrl}
\hline
\noalign{\smallskip}
\multicolumn{1}{c}{Star}& \multicolumn{1}{c}{Site} &
 Observing period & \multicolumn{1}{c}{N}& \multicolumn{1}{c}{Survey}
& \multicolumn{1}{c}{Reference}\\
 & & & &\multicolumn{1}{c}{[hours]} & \\
\noalign{\smallskip}
\hline
\noalign{\smallskip}
HD 37819  & Merate & Jan. 1986  & 462 & 32 & A\&A 1987, 181, 273\\
\noalign{\smallskip}
HR 1225   &    ESO & Nov. 1987  & 705 & 38& A\&A 1989, 220, 144\\
$o^1$ Eri & ESO    & Nov. 1987  & 710 & 38& A\&A 1989, 220, 144\\
HR 547    &    ESO & Nov. 1987  & 462 & 22& A\&A 1989, 220, 144\\
\noalign{\smallskip}
HD 16439  & Merate & Dec. 1988--Jan. 1989 & 1020 & 54& A\&A 1990, 230, 91 \\
HD 16439  & Merate & Dec. 1994--Jan. 1995 & 854 & 63 & A\&A 1996, 312, 912 \\
\noalign{\smallskip}
HD 101158 &    ESO & Apr. 1989 (phot.) & 1234 &  62   &A\&A 1991, 245, 136 \\
HD 101158 &    ESO & Apr. 1994 (spectr.) &74 & 24&A\&A 1997, 323, 844\\
\noalign{\smallskip}
X Cae     &    ESO & Nov. 1989  & 1013 &    54   &  A\&A 1992, 255, 153\\
X Cae     & ESO & Nov. 1992 & 1634 & 100 & A\&A 1996, 312, 855\\
X Cae     & ESO & Nov. 1992 (spectr.) & 101 & 27& A\&A 1996, 312, 855\\
X Cae     & ESO & Nov. 1996 & 916 & 99 & work in progress\\
X Cae     & ESO & Nov. 1996 (spectr.) &  & & work in progress\\
\noalign{\smallskip}
44 Tau    & Merate & Dec. 1989--Feb. 1990  & 2434 & 117&  A\&A 1992, 256, 113\\
\noalign{\smallskip}
BI CMi&    ESO & Jan.--Feb. 1991 & 1390 & 100   &  A\&A 1994, 281, 66 \\
BI CMi&  Merate & Jan.--Feb. 1991 &  642 &  43   &  A\&A 1994, 281, 66\\
\noalign{\smallskip}
HD 224639& ESO & Sept.--Oct. 1991  & 2567 & 120  & A\&A 1995, 299, 427 \\
HD 224639& ESO & Oct.--Nov. 1994  & 1917 & 71    & A\&A 1996, 308, 847 \\
HD 224639& Multisite & Sept.-Oct. 1995  &1252  &125  & work in progress \\
\noalign{\smallskip}
HD 18878  & Merate & Nov. 1991--Jan. 1992 & 2915 & 150 & A\&A 1993, 274, 811 \\
HD 19279  & Merate & Nov. 1991--Jan. 1992 & 2679 & 150 & A\&A 1993, 274, 811 \\
\noalign{\smallskip}
$\beta$ Cas & Merate & Oct. 1986; Oct. 1992 & 925 & 30  &
                                             A\&AS 1994, 108, 55\\
$\beta$ Cas & Asiago & Dec. 1993 (spectr.) & &  & work in progress \\
\noalign{\smallskip}
FG Vir  & ESO & Apr.--May 1992 & 792 & 53 & A\&A 1994, 287, 95 \\
FG Vir  & ESO & Apr.--May 1992 (spectr.) & 22 & 8 & A\&A 1994, 287, 95 \\
FG Vir  & Multisite & Apr.--May 1995 &435  &315 & A\&A 1998, 331, 271  \\
\noalign{\smallskip}
BF Phe & ESO & Sept. 1993 & 832 & 79  & A\&A 1996, 312, 912 \\
\noalign{\smallskip}
AZ CMi  & Merate & Jan.--Feb. 1994 & 307 & 28  & A\&A 1996, 312, 912 \\
\noalign{\smallskip}
HD 2724& ESO & Sept.-Oct. 1993 &884  &90  & A\&A 1998, 336, 518 \\
HD 2724& ESO & Sept.-Oct. 1993 (spectr.) &154  &34  & A\&A 1998, 336, 518 \\
HD 2724& ESO & Oct. 1997 (spectr.) &189  &52  & work in progress \\
\noalign{\smallskip}
BV Cir  & ESO & June 1996 &  &  & work in progress \\
BV Cir  & ESO & June 1996 (spectr.) & 118 & 38 & work in progress \\
BV Cir  & ESO & June 1998 (spectr.) & 156 & 48  & work in progress \\
\hline
\end{tabular}
\end{table}

When the star is multiperiodic, it is
very complicated to make out the power spectrum owing to the interaction
between the excited modes and the spectral window. If measurements are performed
from a single site, not only the  excited frequency $f_1$ will be detected,
but also its aliases $f_1\pm n$ (where $n$ is an integer number of c~d$^{-1}$).
When several
frequencies are simultaneously excited, it is not an easy task to correctly
separate the true frequencies from the aliases owing to the large number of
peaks visible in the power spectra. To simplify the analysis and to proceed
to an identification of a large number of excited modes
(i.e. to do {\it asteroseismology}), it is necessary to
deal with a better spectral window and this can be ensured by multisite
observations, i.e. carried out at different longitudes; we see above the
case of FG Vir. 

The key problem is to typify the modes, i.e. to assign the two $\ell, m$
numbers to each detected frequency. From a photometric point of view,
this can be done by calculating the
phase shifts between light curves in different colours ($UBV$ and
$uvby$). However, the detection of these shifts is a very delicate
procedure and observational errors can hamper a reliable determination.

\subsection{The spectroscopic approach}

An independent promising  approach to mode identifications
is offered by the study of line profile variations  in high
resolution spectrograms.  Actually,  such variability has been
detected in $\delta$ Sct stars and powerful techniques for their
analysis have been developed.
On the basis of spectroscopic data, it is possible to calculate the line
moments (Balona 1987; Aerts at al. 1992) and, from the amplitudes and
phases of their
variations during a pulsation cycle, to estimate
\begin{itemize}
\item the degree $\ell$ and the  azimuthal number $m$ of each pulsation mode;
\item the amplitudes and the phases of the velocity fields and temperature
 variations;
\item the inclination of the rotational axis
and the equatorial rotational velocity.
\end{itemize}

This technique is very suitable for low--degree modes and it has been
already applied by us to some $\delta$ Sct stars (FG Vir and X Cae),
but its full exploitation has been hampered by the lack of an adequate frequency
resolution of the moment curves, caused by the too short baseline
of observations. Independent constraints on mode
typifying and detection of higher degree $\ell$ modes can be obtained from the
analysis of the time series of the individual pixels defining the line
profiles (Telting \& Schrijvers 1997, Mantegazza 1997).

\subsection{The synergic approach}

Trying to fill the gap between the outputs of the observations and the
inputs necessary for a reliable pulsational model,
Bossi et al. (1998) discriminated between different modes through a
direct fit of pulsational model
to spectroscopic and photometric data, i.e. by applying the
 {\it synergic approach} 
(see also Zerbi, this volume, for a more detailed discussion).
Its importance is stressed by Fig.~6: the two techniques are able to 
detect different modes excited in the stellar atmosphere of $\delta$ Sct
variables and only by
combining them we can have a full description of the pulsational content.

\begin{figure*}
\centerline{\psfig{figure=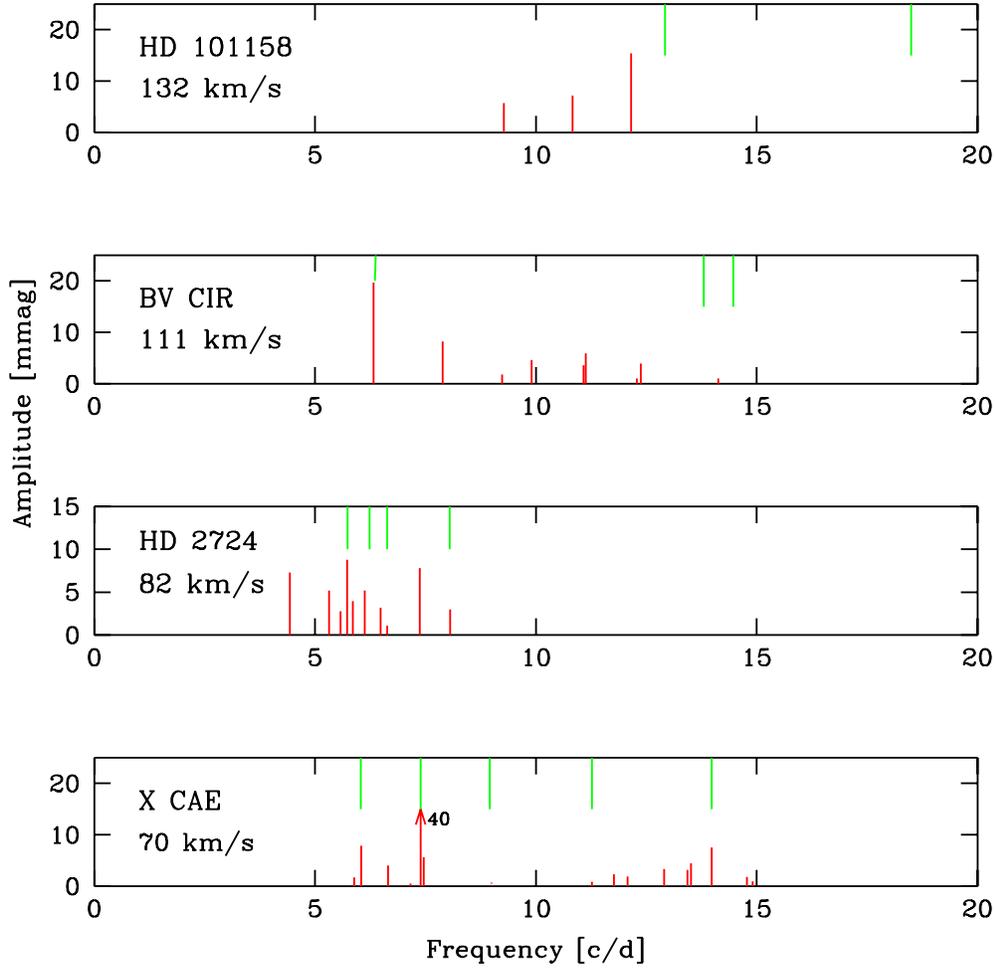,width=14.cm}}
\caption{The frequencies detected photometrically (in red, from bottom
to top) and spectroscopically (in green, from top to bottom) in a sample
of $\delta$ Sct star. $v \sin i$ values are also reported. Note that the
two different techniques detected different sets of excited modes.}
\end{figure*}

\begin{figure*}
\centerline{\psfig{figure=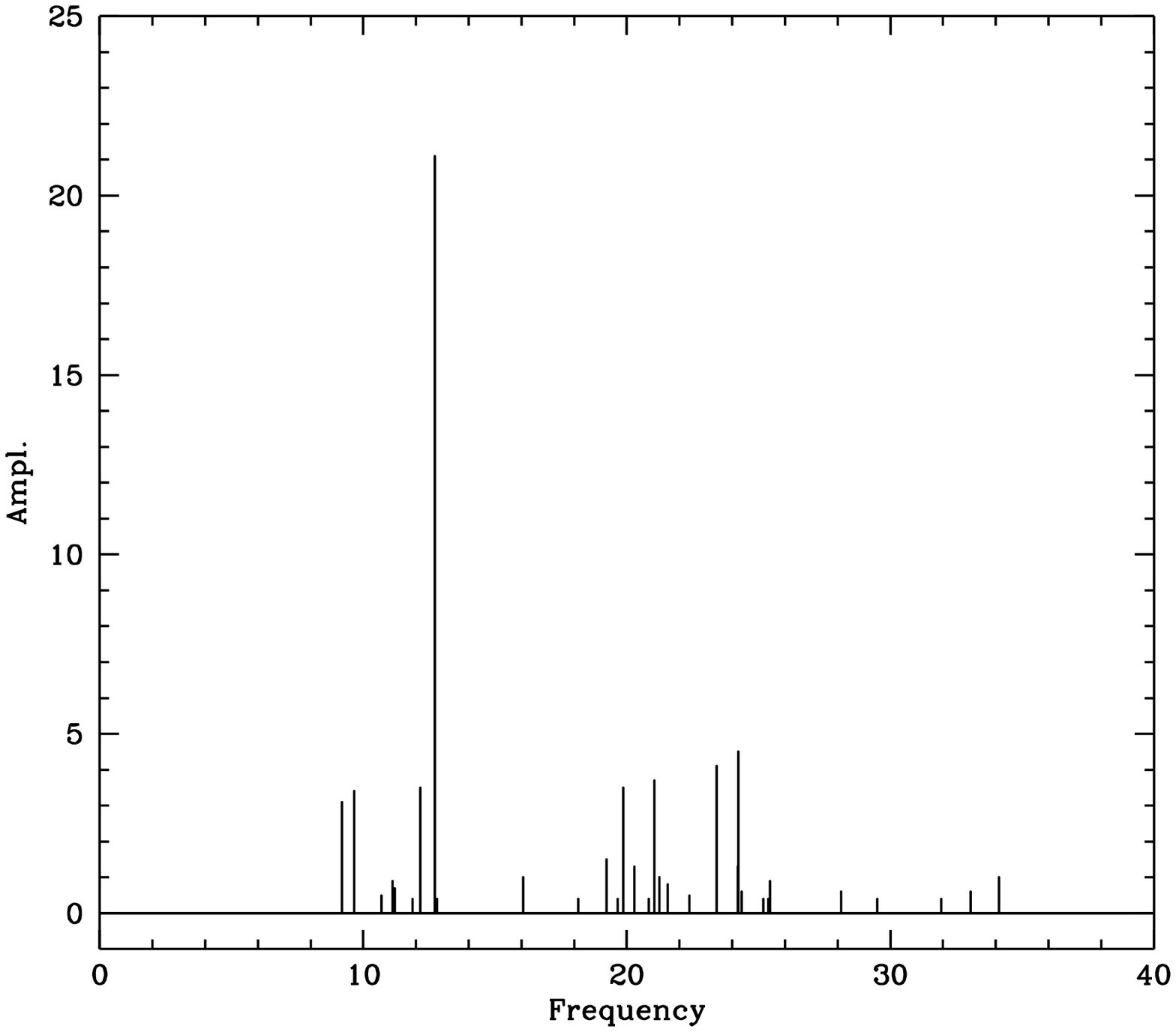,width=14.cm}}
\caption{The frequency content of FG Vir, the best observed and studied
$\delta$ Sct star. These 32 frequencies were detected by means of a 
multisite campaign (Breger et al. 1998). Amplitudes are in mmag,
frequencies in c~d$^{-1}$.}
\end{figure*}

We are trying to increase the
reliability of the mode identification by improving this technique
(these works are marked ``in progress" in Tab.~1).
 As a matter of fact, in the case of HD 2724 (Bossi et al. 1998;  
Mantegazza \& Poretti 1999, in preparation)
it was demonstrated as it is possible to obtain constraints on the inclination
angle; in such a case a much more reliable mode
typifying of the observed modes can be proposed. 

\subsection{FG Vir}

The recent results indicates that, due the very complex phenomenology
of $\delta$ Sct stars (Fig.~6 is again an excellent example), a good way
to proceed is to focalize on the few already well--promising candidates
for asteroseismologic studies.
The rapidly growth in the knowledge of the pulsational content of FG Vir
is an evident example: Mantegazza et al. (1994) discovered 7 pulsation
modes by means of 792 $V$ measurements in 8 nights; Breger et al. (1995)
pushed this boundary to 10 frequencies (the first 7 being confirmed)
by means of a multisite campaign covering 170 hours; more recently, Breger
et al. (1998) secured 435 hours at 6 observatories during a
time span of 40 days, extracting up to 32 frequencies in its light curve
(Fig.~7 shows their spacings and amplitudes).
From a  spectroscopic point of view, Viksum et al. (1998)
determinated the ratios between equivalent widths and
photometric amplitudes,
assigning $\ell$ values to the 8 strongest modes, but no $m$ value could be
proposed. This identification was confirmed by Breger et al. (1999),
who compared the observed photometric phase lags between $v$
and $y$ light curves with the theoretical values for different $\ell$.
Theoretical models of FG Vir were also proposed by Guzik et al. (1998);
results are satisfactory, but the
authors stressed how important it is to propose more secure mode
identifications. This is another confirmation of the necessity of
the synergic approach.

\section{The $\gamma$ Dor stars}

\subsection{The multiperiodic content of the light curves}

In recent years few early F--type stars showing small amplitude light variations
(a few hundredths of a mag) have been discovered, mainly because these stars
were used as comparison stars to measure $\delta$ Sct variables. The
periods are  much longer than those observed in $\delta$ Sct stars; since
multiperiodicity is currently observed, $g$~--~mode pulsations are
 the most plausible cause of the
variability.  The existence of a such a class of pulsating variable
stars of spectral type $\sim$F0 is now accepted and $\gamma$ Dor has been
designated as its prototype. Zerbi \& Kaye (this volume) and Kaye et al.
(1999) provide an exhaustive discussion of their phenomenology and a 
rigorous classification of their kind of variability. The reader can find
in Kaye et al. (1999) the reference papers on all the {\it bona fide}
members of the class, included  the stars quoted below.

\begin{table}
\caption{Absolute predicted differences between the frequency $f_o$ in
the rest-frame of the star and the frequency $f$ for $|m|=1$ in the cases
$\ell=1$ and $\ell$=2. Cases 1 and 2 are extreme $\sin i$ conditions, i.e.
$\sin i = 0.1$ and $\sin i = 0.9$, respectively. $\Omega$ is the rotational
frequency and $<f-f_o>$ values are the mean observed spacings.}
\begin{tabular}{cccccccc}
\hline
Star&Case& $v\sin i$ & $R$ & $\Omega$ & $|f-f_o|$ & $|f-f_o|$ & $<|f-f_o|>$\\
 & & [Km~s$^{-1}$] & $[R_\odot]$ & [d$^{-1}$] & $\ell=1$ & $\ell$=2& Observed \\
\hline
HD 224945  &1& 55.00& 1.43& 4.77& 2.39& 3.98& 0.17-1.84 \\
      &2&   &   & 0.53& 0.27& 0.44& \\
$\gamma$ Dor &1& 62.00& 1.70& 4.53& 2.26& 3.77& 0.04-0.15\\
      &2&   &   & 0.50& 0.25& 0.42& \\
9 Aur    &1& 18.00& 1.62& 1.38& 0.69& 1.15& 0.03-0.45\\
      &2&   &   & 0.15& 0.08& 0.13& \\
BS 2740   &1& 40.00& 1.64& 3.03& 1.51& 2.52& 0.04-0.14\\
      &2&   &   & 0.34& 0.17& 0.28& \\
HD 62454  &1& 53.00& 2.02& 3.26& 1.63& 2.71& 0.14-0.30\\
      &2&   &   & 0.36& 0.18& 0.30& \\
HD 108100  &1& 68.00& 2.01& 4.20& 2.10& 3.50& 0.08\\
      &2&   &   & 0.47& 0.23& 0.39& \\
HR 8799   &1& 45.00& 1.46& 3.83& 1.91& 3.19& 0.08-0.33\\
      &2&   &   & 0.43& 0.21& 0.35& \\
\hline
\end{tabular}
\end{table}

There are two other differences between $p$-- and $g$--mode pulsations,
which can be considered as fingerprints for the latter:
\begin{itemize}
\item when considering the $(b-y)$--$c_1$ diagrams the
$\delta$ Sct stars (i.e. the  $p$--mode pulsators) describe ellipses,
while $\gamma$ Dor loops are only a little
deviating from a straight line; 
\item Mantegazza et al. (1995) stressed the
phase correlation between light and radial velocity curves, i.e. the
opposite behaviour showed by radial and nonradial $p$--mode pulsation;
\end{itemize}

Let me discuss here the characteristics of the multiperiodicity observed
in $\gamma$ Dor stars.
In HR 2740, HD 108100 and HD 62454 the periods are very close
each other, while in HD 224945 they are spread in a larger interval.
In $\gamma$ Dor two periods very stable in amplitude and in phase were
observed for  a long time, but a third one appeared in the last decade. This
is not an isolated case, since in the 9 Aur light curve the 1.375 d
period appeared, while the 1.302 d  disappeared. In HD 224945
the amplitudes of the excited modes are continuously changing, not only
as absolute value, but also modifying the internal ranking between them;
moreover, some periods are not always observed.

These characteristics and in general  such a large variety of behaviours
remind us the phenomenology of $\delta$ Sct stars. On the other hand,
the change of the shape of some light curve around maxima
(HD 164615 and HR 8330 are the two stars where this fact is more evident)
is a peculiarity of $\gamma$~Dor stars and it should 
be related straightly to the $g$--modes excitation mechanism.

\subsection{Can the rotational splitting explain the observed
multiperiodicity~?}

At present, the fact that the variability of the 50\% of the $\gamma$ Dor
stars can be explained by a single period (even if  modulated in amplitude,
as in the case of HD~164615) is another difference respect with the $\delta$
Sct stars, where monoperiodicity is uncommon in small amplitude ($<$ 0.10
mag) objects. The question if it is
possible to explain the rich power spectra as an effect of
a rotational splitting, i.e. modes with
same $\ell$ but different $m$, raises in a natural way.
In this case, a predictable relation between
observed frequencies should exist.
For the purposes of this discussion, we will use the approach described 
by Aerts and Krisciunas (1996) for first order rotational splitting:
\begin{equation}
|f - f_{0}|= m\Omega\left(1-{{1}\over{\ell(\ell+1)}}\right).
\end{equation}

Equation (1) provides an estimate of the frequency separation as a
function of the rotational velocity and the spherical degree, $\ell$.
Using a mean value for the disentangled frequencies in each object
and the calculated physical quantities, we determinated the
amount of rotational splitting expected in the multiperiodic objects (Tab.~2).

Among them, the splitting is excluded for the 50\%
of the sample, i.e., $\gamma$ Dor, HD 108100, HR 8799 and HR 2740.
Therefore, even if rotational splitting cannot be excluded
for individual objects and the theoretical approach  needs an
improvement, it is hard to admit it as a cause of the multiperiodic behaviour
observed in the class as a whole.

Also in the case of $\gamma$ Dor stars spectroscopy can supply a
different approach to the the study of $g$--mode pulsations, but such
observations were performed for a few stars (mainly 9 Aur and $\gamma$ Dor
itself) and more efforts are necessary to obtain a clear picture. However,
the proposed typifyings seem to rule out the action of the rotational
splitting.

As a final remark, the two $\gamma$ Doradus
identified in double stars (HD 62863$\equiv$2 Puppis B and HD 62454) can
be considered a good target to verify the connection between duplicity and
pulsation and in what way the tidal effects can act
(see Harmanec et al. 1997 for a similar approach in studying $Be$ stars).

\section{Acknowledgements}

The author wishes to thank A.~Kaye, L.~Mantegazza and F.M.~Zerbi for
useful discussions.

\section{Question}

J.~CHRISTENSEN-DALSGAARD -- In relation to the problems of data analysis and
the use of standard techniques as {\sc clean}, I wonder whether tests have
been made by exchanging data, or by applying the techniques in blind
tests using synthetic data~?
\parindent=0truecm

E.~PORETTI -- Exchanges of data are very common between researchers in our
field. The same sets of data are often re-analyzed in subsequent works.
Our group always performs different data analysis by using different
techniques: iterative sine--wave fitting, {\sc clean}, Fourier transform~...
An example  is constituted by the analysis of HD~224639 (Mantegazza et al. 
1995, A\&A 299, 437; Mantegazza et al. 1996, A\&A 308, 847). Synthetic
data are also currently generated to evaluate the effects of noise.

\end{document}